\documentclass[ALICE,manyauthors]{cernphprep}
\usepackage[comma,square,numbers,sort&compress]{natbib}
\usepackage{hyperref}
\usepackage{lineno}
\usepackage{xspace}
\usepackage[T1]{fontenc}
\usepackage{orcidlink}

\begin{document}
%

\newcommand{\pp}           {pp\xspace}
\newcommand{\ep}           {ep\xspace}
\newcommand{\ppbar}        {\mbox{$\mathrm {p\overline{p}}$}\xspace}
\newcommand{\XeXe}         {\mbox{Xe--Xe}\xspace}
\newcommand{\PbPb}         {\mbox{Pb--Pb}\xspace}
\newcommand{\pA}           {\mbox{pA}\xspace}
\newcommand{\pPb}          {\mbox{p--Pb}\xspace}
\newcommand{\AuAu}         {\mbox{Au--Au}\xspace}
\newcommand{\dAu}          {\mbox{d--Au}\xspace}

\newcommand{\s}            {\ensuremath{\sqrt{s}}\xspace}
\newcommand{\snn}          {\ensuremath{\sqrt{s_{\mathrm{NN}}}}\xspace}
\newcommand{\pt}           {\ensuremath{p_{\rm T}}\xspace}
\newcommand{\meanpt}       {$\langle p_{\mathrm{T}}\rangle$\xspace}
\newcommand{\ycms}         {\ensuremath{y_{\rm CMS}}\xspace}
\newcommand{\ylab}         {\ensuremath{y_{\rm lab}}\xspace}
\newcommand{\etarange}[1]  {\mbox{$\left | \eta \right |~<~#1$}}
\newcommand{\yrange}[1]    {\mbox{$\left | y \right |~<~#1$}}
\newcommand{\dndy}         {\ensuremath{\mathrm{d}N_\mathrm{ch}/\mathrm{d}y}\xspace}
\newcommand{\dndeta}       {\ensuremath{\mathrm{d}N_\mathrm{ch}/\mathrm{d}\eta}\xspace}
\newcommand{\avdndeta}     {\ensuremath{\langle\dndeta\rangle}\xspace}
\newcommand{\dNdy}         {\ensuremath{\mathrm{d}N_\mathrm{ch}/\mathrm{d}y}\xspace}
\newcommand{\Npart}        {\ensuremath{N_\mathrm{part}}\xspace}
\newcommand{\Ncoll}        {\ensuremath{N_\mathrm{coll}}\xspace}
\newcommand{\dEdx}         {\ensuremath{\textrm{d}E/\textrm{d}x}\xspace}
\newcommand{\RpPb}         {\ensuremath{R_{\rm pPb}}\xspace}

\newcommand{\nineH}        {$\sqrt{s}~=~0.9$~Te\kern-.1emV\xspace}
\newcommand{\seven}        {$\sqrt{s}~=~7$~Te\kern-.1emV\xspace}
\newcommand{\twoH}         {$\sqrt{s}~=~0.2$~Te\kern-.1emV\xspace}
\newcommand{\twosevensix}  {$\sqrt{s}~=~2.76$~Te\kern-.1emV\xspace}
\newcommand{\five}         {$\sqrt{s}~=~5.02$~Te\kern-.1emV\xspace}
\newcommand{\thirteen}         {$\sqrt{s}~=~13$~Te\kern-.1emV\xspace}
\newcommand{\twosevensixnn}{$\sqrt{s_{\mathrm{NN}}}~=~2.76$~Te\kern-.1emV\xspace}
\newcommand{\fivenn}       {$\sqrt{s_{\mathrm{NN}}}~=~5.02$~Te\kern-.1emV\xspace}
\newcommand{\eightnn}       {$\sqrt{s_{\mathrm{NN}}}~=~8.16$~Te\kern-.1emV\xspace}
\newcommand{\LT}           {L{\'e}vy-Tsallis\xspace}
\newcommand{\GeVc}         {Ge\kern-.1emV/$c$\xspace}
\newcommand{\MeVc}         {Me\kern-.1emV/$c$\xspace}
\newcommand{\TeV}          {Te\kern-.1emV\xspace}
\newcommand{\GeV}          {Ge\kern-.1emV\xspace}
\newcommand{\MeV}          {Me\kern-.1emV\xspace}
\newcommand{\GeVmass}      {Ge\kern-.1emV/$c^2$\xspace}
\newcommand{\MeVmass}      {Me\kern-.1emV/$c^2$\xspace}
\newcommand{\lumi}         {\ensuremath{\mathcal{L}}\xspace}

\newcommand{\ITS}          {\rm{ITS}\xspace}
\newcommand{\TOF}          {\rm{TOF}\xspace}
\newcommand{\ZDC}          {\rm{ZDC}\xspace}
\newcommand{\ZDCs}         {\rm{ZDCs}\xspace}
\newcommand{\ZNA}          {\rm{ZNA}\xspace}
\newcommand{\ZNC}          {\rm{ZNC}\xspace}
\newcommand{\SPD}          {\rm{SPD}\xspace}
\newcommand{\SDD}          {\rm{SDD}\xspace}
\newcommand{\SSD}          {\rm{SSD}\xspace}
\newcommand{\TPC}          {\rm{TPC}\xspace}
\newcommand{\TRD}          {\rm{TRD}\xspace}
\newcommand{\VZERO}        {\rm{V0}\xspace}
\newcommand{\VZEROA}       {\rm{V0A}\xspace}
\newcommand{\VZEROC}       {\rm{V0C}\xspace}
\newcommand{\Vdecay} 	   {\ensuremath{V^{0}}\xspace}

\newcommand{\ee}           {\ensuremath{\rm e^{+}e^{-}}\xspace} 
\newcommand{\bb}           {\ensuremath{\mathrm{b\bar{b}}}\xspace}
\newcommand{\cc}           {\ensuremath{\mathrm{c\bar{c}}}\xspace} 
\newcommand{\pip}          {\ensuremath{\pi^{+}}\xspace}
\newcommand{\pim}          {\ensuremath{\pi^{-}}\xspace}
\newcommand{\kap}          {\ensuremath{\rm{K}^{+}}\xspace}
\newcommand{\kam}          {\ensuremath{\rm{K}^{-}}\xspace}
\newcommand{\pbar}         {\ensuremath{\rm\overline{p}}\xspace}
\newcommand{\kzero}        {\ensuremath{{\rm K}^{0}_{\rm{S}}}\xspace}
\newcommand{\lmb}          {\ensuremath{\Lambda}\xspace}
\newcommand{\almb}         {\ensuremath{\overline{\Lambda}}\xspace}
\newcommand{\Om}           {\ensuremath{\Omega^-}\xspace}
\newcommand{\Mo}           {\ensuremath{\overline{\Omega}^+}\xspace}
\newcommand{\X}            {\ensuremath{\Xi^-}\xspace}
\newcommand{\Ix}           {\ensuremath{\overline{\Xi}^+}\xspace}
\newcommand{\Xis}          {\ensuremath{\Xi^{\pm}}\xspace}
\newcommand{\Oms}          {\ensuremath{\Omega^{\pm}}\xspace}
\newcommand{\degree}       {\ensuremath{^{\rm o}}\xspace}
\newcommand{\jpsi}         {\ensuremath{\mathrm{J/\psi}}\xspace}
\newcommand{\psis}         {\ensuremath{\mathrm{\psi(2S)}}\xspace}
\newcommand{\mee}         {\ensuremath{m_{\mathrm{ee}}}\xspace}
\newcommand{\ptee}           {\ensuremath{p_{\rm T}^{\mathrm{ee}}}\xspace}

\begin{titlepage}
\PHyear{2026}       
\PHnumber{042}      
\PHdate{17 February}  
\title{Inclusive \psis production at midrapidity in \pp collisions at \s = 13 TeV}
\ShortTitle{Inclusive \psis production at midrapidity}   

\Collaboration{ALICE Collaboration\thanks{See Appendix~\ref{app:collab} for the list of collaboration members}}
\ShortAuthor{ALICE Collaboration} 

\begin{abstract}
Inclusive production of \psis mesons is measured for the first time at midrapidity ($|y| < 0.9$) in pp collisions at \s = 13 TeV using the ALICE detector at the LHC. The measurement is performed in the transverse-momentum (\pt) range of 4 to 16~\GeVc and is based on events triggered by the Transition Radiation Detector, corresponding to an integrated luminosity of $1.7 \pm 1.8\%$ (syst.)~pb$^{-1}$. The \pt-differential production cross section of the \psis and its ratio to the \jpsi production cross section are presented. The results extend the accessible \pt range down to 4~\GeVc, and a mild rise of the \psis-to-\jpsi cross section ratio with increasing \pt is observed. The measurements are compared with theoretical predictions based on NRQCD and the ICEM model. 
\end{abstract}
\end{titlepage}

\setcounter{page}{2} 


\section{Introduction}
Charmonia, such as \jpsi or \psis, are bound states of a valence charm and anti-charm quark. Charmonia can be produced directly in the hadronic interaction, from decays of higher-mass charmonium states, or from weak decays of beauty hadrons. The first two contributions are referred to as prompt production, while the latter corresponds to non-prompt production. At low transverse momenta, direct production dominates, while beyond transverse momenta of 10~\GeVc, non-prompt production from decays of beauty hadrons becomes comparable to the direct component~\cite{LHCb:2015foc,ALICE:2017leg}. 

The prompt production is governed by different scales, in particular the hard scale of the charm-quark mass and the soft scales of the charm-quark momenta in the charmonium rest frame and the charmonium binding energy~\cite{nrqcd}. Because of these very different scales, theoretical calculations assume that the hard production of the \cc pair and the soft hadronization process factorize~\cite{Brambilla:2010cs}.
In this ansatz, one separates the short-distance perturbative hard scattering process from the long-distance non-perturbative dynamics, which are encoded in long-distance matrix elements (LDMEs). The dominant process for the production of a \cc pair at LHC energies is gluon fusion. The LDMEs, obtained from global fits to a wide range of quarkonium production and polarization measurements, are commonly assumed to be universal.
However, recent studies of open heavy-flavor fragmentation suggest that universality in heavy-flavor hadronization, i.e. independence of the collision system and energy, does not hold: the ALICE Collaboration observed significantly enhanced production of charmed baryons compared to results from electron–positron and electron–proton collisions~\cite{ALICE:2021dhb}, and a similar enhancement was recently reported by the LHCb Collaboration in the beauty sector~\cite{LHCb:2023wbo}. While fragmentation fractions and LDMEs describe different aspects of non-perturbative dynamics, these findings highlight that heavy-flavor hadronization can depend on the collision environment. Comparatively, much less is known experimentally about hadronization of charm-quark pairs into quarkonia.

Different treatments of the non-perturbative processes have led to various theoretical models of charmonium formation. Shortly after the discovery of the \jpsi, the Color Singlet Model (CSM)~\cite{Chang:1979nn,Baier:1981uk,Berger:1980ni,Einhorn:1975ua,Carlson:1976cd} was introduced. Under the assumption that the \cc pair evolving into the quarkonium state is in a color-singlet state (CS), the production of on-shell heavy-quark pairs with the same quantum numbers as the quarkonium state is modeled using perturbative Quantum Chromo Dynamics (pQCD). The production rates are determined by the CS \cc wave functions and their derivatives and can be related to quarkonium decay rates. While successful at low energies, this model significantly underestimates the observed production cross section, even using calculations at next-to-leading order (NLO) in the strong coupling constant $\alpha_s$~\cite{Lansberg:2010vq} at the Tevatron and LHC energies~\cite{CDF:1997ykw,ALICE:2014uja}. The Non-Relativistic Quantum ChromoDynamics framework (NRQCD)\cite{Bodwin:1994jh} extends the CSM by including contributions from heavy-quark pairs produced in a color-octet (CO) state, in addition to the CS contributions. The transition probability from the CO \cc state to a CS charmonium state is expanded in powers of the relative velocity between the heavy quarks and parameterized using LDMEs. These LDMEs are free parameters fitted to experimental data. Recently, NLO extensions of NRQCD~\cite{Ma:2010jj,Butenschoen:2010rq,Xu:2012am} have successfully described quarkonium production cross sections for many different collision systems and energies. At the same time, they struggle to account for the measured polarization at intermediate and high \pt from the CDF Collaboration at the Tevatron, while ALICE data at the LHC at intermediate \pt are well reproduced~\cite{Chao:2012iv,Butenschoen:2012px,Ma:2018qvc,Butenschoen:2022qka}. 

The Color Evaporation Model (CEM)~\cite{Halzen:1977rs,Fritzsch:1977ay,Amundson:1996qr} is based on the concept of quark-hadron duality. It is assumed that the heavy-quark pair evolves into a quarkonium state when the invariant mass is below the production threshold corresponding to twice the mass of the lightest open heavy-flavor meson. The probability of the quark pair to evolve into a specific quarkonium state is determined by one proportionality factor that is assumed to be universal. Once the factor is fixed, transverse-momentum and rapidity ($y$) distributions of the quarkonium can be predicted without additional free parameters. Consequently, the ratio between the yields of two quarkonium states formed from a given heavy-quark pair is expected to be independent of collision energy, \pt, and $y$. The Improved Color Evaporation Model (ICEM)~\cite{Ma:2016exq} employs the $k_{\mathrm T}$-factorization approach, which introduces a \pt dependence due to the \pt of incoming gluons and off-shell properties. As in the CEM, the transition probability to a bound charmonium state is proportional to the \cc pair production cross section integrated over an invariant-mass range from the mass of the bound state to twice the mass of the lightest D meson.

Measurements of production rates of different charmonium states in proton--proton (\pp) collisions at the LHC are interesting to test the various production mechanisms. In particular, they provide a testing ground for the concepts of factorization and universal fragmentation. 
Measurements of bottomonium production at the CDF and at the LHC experiments indicated different populations from excited states~\cite{Digal:2001ue,Boyd:2023ybk}. This suggests that similar variations may arise in charmonium production because charmonium states are more sensitive to higher-order QCD corrections~\cite{Chen:2013itc} and to medium effects in heavy-ion collisions~\cite{Lee:2000csl}. The charmonium measurements in \pp collisions also provide a baseline for proton-nucleus and nucleus--nucleus results, where charmonia are prominent observables to study the quark--gluon plasma~\cite{Andronic:2025jbp}.

In this paper, we report results on the inclusive \pt-differential production cross sections of \psis at midrapidity ($|y| < 0.9$) and the \psis-to-\jpsi cross section ratio measured in \pp collisions at \thirteen for the first time. While the inclusive \jpsi production cross section in \pp collisions has previously been published using minimum-bias (MB) triggered data~\cite{ALICE:2021dtt}, the ratio presented here is obtained consistently within the present analysis. This was made possible by using a level-1 trigger of the Transition Radiation Detector (\TRD) to enhance the sampled luminosity. 

These results complement the measurement at forward rapidity by the ALICE Collaboration~\cite{ALICE:2017leg} and extend the \pt range to lower value compared with midrapidity measurements by the CMS Collaboration for $\pt > 20$~\GeVc~\cite{CMS:2017dju} 
and by the ATLAS Collaboration for $\pt > 8$~\GeVc~\cite{ATLAS:2023qnh}. Measurements of the \psis production cross section in \pp collisions at midrapidity were already performed by the CMS and ATLAS Collaborations for $\pt > 6.5$~\GeVc and $\pt > 10$~\GeVc, respectively, but at lower collision energies, as reported in Refs.~\cite{CMS:2011rxs,ATLAS:2016kwu,CMS:2018gbb}.

In the following, the ALICE detector, the trigger setup, and the data sample are described in Section~\ref{setup}. The data analysis and the determination of systematic uncertainties are detailed in Section~\ref{analysis}. The results are presented and discussed in the context of recent model calculations in Section~\ref{results}, and the summary is given in Section~\ref{summary}.

\section{Experimental setup, data sample, and event selection}\label{setup}
This section focuses on the detectors used for reconstructing the \jpsi and \psis at midrapidity. A comprehensive description of the ALICE detector and its performance can be found in Refs.~\cite{ALICE:2008ngc,ALICE:2014sbx}.

Electron and positron tracks from \jpsi and \psis decays are reconstructed as global tracks using the Inner Tracking System (\ITS)~\cite{ALICE:2010tia} and the Time Projection Chamber (\TPC)~\cite{Alme:2010ke}, which cover the full azimuthal angle and the pseudorapidity interval $|\eta| < 0.9$. 
These detectors are placed within a solenoid magnet providing a uniform magnetic field of $B = 0.5 \, \mathrm{T}$ parallel to the beam direction. The ITS is composed of six layers of detectors. From the innermost to the outermost layers, it includes two layers each of silicon pixel, drift, and strip detectors, allowing for precise reconstruction of primary and secondary vertices. The Silicon Pixel Detector~(\SPD) layers, are placed at radial distances of 3.9 and 7.6 cm from the nominal Interaction Point~(IP) and are used for reconstructing the primary vertex, providing crucial information for track reconstruction.
The \TPC is a cylindrical gas drift detector located at radial distances between 85 and 250 cm from the IP and covers $\pm$ 250 cm from the IP along the beam direction.
It provides particle-identification (PID) information through the measurement of specific ionisation energy loss (\dEdx), as well as up to 159 space points for reconstructing charged-particle tracks.
The Transition Radiation Detector~(\TRD)~\cite{Acharya:2017lco} consists of 522 drift chambers arranged in six concentric layers at radial distances between 2.90 and 3.68 m from the IP, in five stacks along the longitudinal direction, covering the pseudorapidity interval $|\eta| < 0.84$. Each chamber contains a foam/fibre sandwich radiator, followed by a drift region and a multiwire proportional chamber (MWPC) with an amplification and readout section, both filled with a Xe-CO$_2$ gas mixture. The \TRD provides three-dimensional position information, used to reconstruct short local track segments (tracklets) in each chamber.
These tracklets are then used for track reconstruction and for electron identification through the measured charge deposition and, in particular, through transition radiation. The V0 counters~\cite{ALICE:2013axi}, two arrays of 32 scintillator tiles each, have a pseudorapidity coverage of $2.8 < \eta < 5.1$ (V0A) and $-3.7 < \eta < -1.7$ (V0C) and provide trigger and event-selection information.

The data used in this analysis have been recorded with the \TRD trigger by the ALICE Collaboration during the LHC Run 2 data-taking campaigns of 2017 and 2018 for \pp collisions at \thirteen. A MB data sample, triggered by the coincidence of signals in the V0A and V0C detectors and collected during the same campaign, is used to estimate the \TRD-trigger efficiency. Events are required to have a reconstructed vertex located within $|{z_{\mathrm{vtx}}}| < 10$ cm from the IP along the beam direction. The SPD and V0 detectors are used to reject background from beam-gas interactions and pile-up events using algorithms that identify multiple vertices, as described in Ref.~\cite{ALICE:2014sbx}. The remaining fraction of pile-up events is negligible. 

For these campaigns, a \TRD trigger at level 1, about 6 $\mu$s after the collision was introduced to trigger on electrons and positrons of intermediate and high \pt.
A single-electron trigger with a \pt threshold of 2~\GeVc and likelihood selections based on the \TRD signal was employed to increase the probability of selecting events containing quarkonia decays. This trigger configuration substantially enhances the recorded yields of inclusive \jpsi and \psis candidates compared with the MB trigger. An enhancement factor is defined as the ratio of the number of reconstructed electron candidates per event in the \TRD-triggered sample to that in the MB sample. This factor is approximately 34 for the present \pp dataset at \thirteen, illustrating the enrichment of electron-containing events provided by the \TRD trigger.
Additional track quality criteria are imposed, requiring five tracklets (reconstructed in individual TRD layers) per stack, including a tracklet in the first layer, and a minimum value for the TRD track sagitta ($\Delta p_{\mathrm{T}}^{-1} < 0.2~c/\mathrm{GeV}$) of the TRD track to reduce background from photon conversions in the detector material before the TRD. For more details on the use of the \TRD trigger for \jpsi measurements, see Ref.~\cite{ALICE:2022zig}.

The total number of events after selection ($N_{\rm TRD}$) is $1.78 \times 10^{8}$, corresponding to an integrated luminosity of $1.7\,\mathrm{pb}^{-1} \pm 1.8\%$ (syst.). The integrated luminosity was evaluated separately for each year based on the MB trigger cross section determined from van der Meer scans~\cite{vanderMeer:1968zz}, as described in Ref.~\cite{ALICE:2021leo}.

\section{Data analysis}\label{analysis}
Inclusive \jpsi and \psis meson production is measured using \TRD-triggered events, similar to the measurement of \jpsi production in p--Pb collisions in ALICE~\cite{ALICE:2022zig}. Both \jpsi and \psis mesons are reconstructed via the dielectron decay channels, with branching ratios (BR) of $5.971 \pm 0.032\%$ and  $0.794 \pm 0.022\%$ for \jpsi and \psis, respectively~\cite{pdg}. Due to the trigger condition requiring a single electron with transverse momentum of \pt $\ge 2~ $\GeVc, the acceptance and efficiency for \jpsi and \psis mesons are limited to $\pt > 2$~\GeVc.

The \pt-differential production cross section is obtained by correcting the raw number of reconstructed charmonium candidates, $N^\mathrm{raw}_{\psi}$, for the acceptance and efficiency ($A \times \epsilon$) in the corresponding \pt and $y$ intervals. The definitions and evaluations of these correction terms are described in Section~\ref{sec:correction}. The full expression for the \pt-differential cross section, including the normalisation to the branching ratio, integrated luminosity, and the widths of the \pt and rapidity intervals, is given by

\begin{equation} 
    \frac{\mathrm{d}^{2}\sigma}{\mathrm{d}y \, \mathrm{d}p_{\mathrm{T}}} = \frac{{N^\mathrm{raw}_{\psi}}}{\left(\it{A} \times \epsilon \right) \times \mathrm{BR} \times \Delta p_{\mathrm{T}} \times \Delta y \times \mathcal{L}_{\mathrm{int}}}.
    \label{eq:xsection1}
\end{equation}

Here, $\mathcal{L}_{\mathrm{int}}$ is the integrated luminosity of the \TRD-triggered data sample given by $\mathcal{L}_{\mathrm{int}} = N_{\mathrm{MB}}/\sigma_{\mathrm{MB}}$, where $\sigma_{\mathrm{MB}}$ is the MB trigger cross section~\cite{ALICE:2021leo}. 
The number of equivalent MB events, $N_{\mathrm{MB}}$, is determined as $N_{\mathrm{MB}} = N_{\mathrm{\TRD}} \times f_{\mathrm{norm}}$, where the number of \TRD-triggered events, $N_{\mathrm{\TRD}}$, is scaled with a normalisation factor, $f_{\mathrm{norm}}$, evaluated for each run. This factor accounts for the enhancement due to the trigger selection and represents the inverse of the probability for a MB event to also satisfy the \TRD-trigger condition. The weighted average value of $f_{\mathrm{norm}}$ across the dataset is approximately 570.

The efficiencies are determined through Monte Carlo (MC) simulations. Minimum-bias \pp events are generated with PYTHIA 6.4 using the Perugia 2011 tune~\cite{py6}, and one \jpsi and one \psis meson are embedded in each event. Prompt \jpsi and \psis are injected with flat rapidity distributions. The \pt distribution of the \jpsi is reweighted to match a measurement at midrapidity at \thirteen~\cite{ALICE:2021dtt}. 
For the \psis, the measurement at forward rapidity at \thirteen~\cite{ALICE:2017leg} is used up to 6~\GeVc, and a flat distribution is assumed at higher \pt. The \pt distributions are then iteratively reweighted following Ref.~\cite{ALICE:2017leg}. In each iteration, the acceptance times efficiency correction is updated using the corrected \pt-differential cross section, and the procedure is repeated until the difference between the input and output corrected spectra is below about 1\%. Both prompt \jpsi and \psis are assumed to be unpolarised. In addition, non-prompt \jpsi and \psis are generated from the decays of beauty hadrons using the PYTHIA generator. Their decays to charmonia are simulated using the EvtGen package~\cite{Lange:2001uf}, which provides a proper description of beauty-hadron decay channels as well as radiative charmonium decays. All generated particles are transported through the ALICE detector modeled using GEANT3~\cite{Brun:1994aa}.

To reproduce the \TRD-trigger condition as in the data, the simulation includes an emulation of the \TRD online trigger logic, applying the electron and positron likelihood estimated from \TRD and track segment requirements used during data taking.
The \TRD-trigger efficiency for electrons or positrons ($\epsilon_{\mathrm{TRD}}(p_{\mathrm{T}}^{\mathrm{e}})$)\footnote{It should be noted that $\epsilon_{\mathrm{TRD}}(p_{\mathrm{T}}^{\mathrm{e}})$ represents the single-track trigger efficiency, whereas $\epsilon_{\mathrm{TRDtrg}}$ discussed later in Section~\ref{sec:correction} denotes the efficiency for a charmonium candidate to pass the trigger requirement via any of its decay electrons.} is defined as
\begin{equation}
    \epsilon_{\mathrm{TRD}}(p_{\mathrm{T}}^{\mathrm{e}}) = \frac{N_{\mathrm{e}}^{\mathrm{TRD,~MB}}(p_{\mathrm{T}}^{\mathrm{e}})}{N_{\mathrm{e}}^{\mathrm{MB}}(p_{\mathrm{T}}^{\mathrm{e}})},
    \label{eq:slEff}
\end{equation}
where $N_{\mathrm{e}}^{\mathrm{MB}}$ is the number of reconstructed single-electron (or positron) tracks identified in MB data using the \TPC, and $N_{\mathrm{e}}^{\mathrm{TRD,~MB}}$ is the number of those tracks that are also selected by the \TRD trigger.

The \TRD-trigger efficiency for electrons and positrons exhibits time-dependent variations due to changes in the magnetic field and the gas mixture, which affect the detector gain. 
Therefore, the \TRD-trigger efficiency is determined by accounting for run-wise dependencies of the magnetic field and gas conditions. This is done by evaluating the efficiency for each run and computing a weighted average using the corresponding fraction of TRD-triggered events~\cite{ALICE:2022zig}. The \pt dependence of the \TRD-trigger efficiency is parametrized with a fit, and the MC simulation is tuned to reproduce the fitted efficiency.

\begin{figure}[htb]
    \centering
        \includegraphics[width=0.98\linewidth]{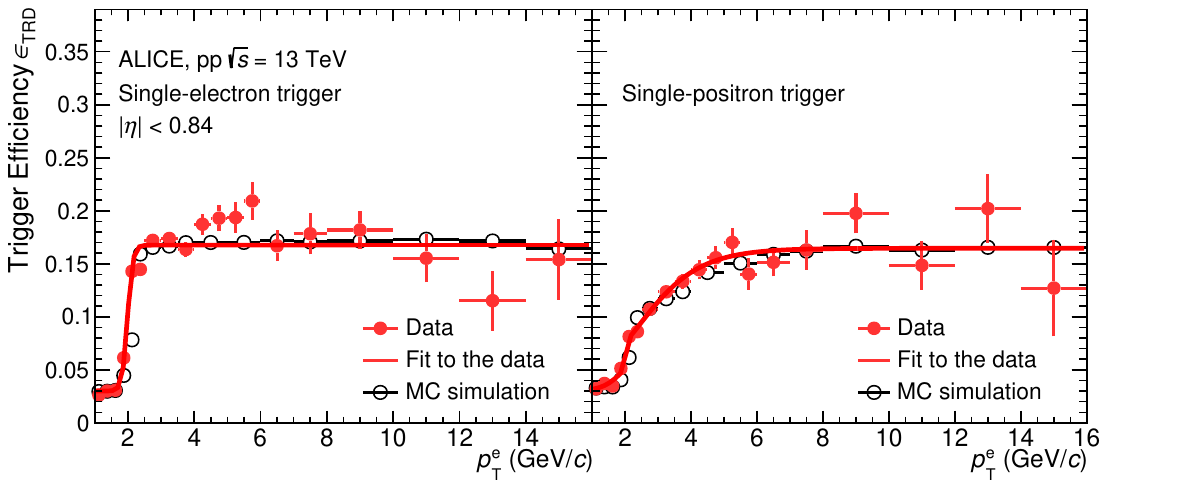}
    \caption{Acceptance times \TRD-trigger efficiency ($\epsilon_{\mathrm{TRD}}(p_{\mathrm{T}}^{\mathrm{e}})$) described by Eq.~\ref{eq:slEff} for single electrons (left) and positrons (right) obtained from \TRD-triggered data (red solid circle), fitted with an error function (red line), and MC simulation (black open circle). See text for details.}
    \label{fig:slEff}
\end{figure}

Figure~\ref{fig:slEff} shows the \TRD-trigger efficiency for electrons (left) and positrons (right) as a function of the reconstructed \pt of the global track. The \pt threshold is clearly visible by the rapid increase in efficiency at 2~\GeVc, followed by a plateau at higher \pt. The plateau value for electrons is slightly higher than the one for positrons and has a sharper turn-on curve. 
Below 2~\GeVc, non-zero values for the \TRD-trigger efficiency for electrons and positrons are found. These non-zero values arise from the presence of high-\pt electrons or positrons in the same event that satisfy the online \TRD-trigger condition, rather than the low-\pt tracks themselves. This contribution is referred to as the underlying-event effect in the following.
An error function is employed to fit the \TRD-trigger efficiency, shown as a red line, with an offset accounting for the underlying-event effect. A softer rise of the trigger efficiency above 2~\GeVc is observed for positrons, which is attributed to the $E \times B$ effect.
This effect originates from the perpendicular orientation of the electric and magnetic fields in the drift region of the \TRD chambers, which causes drifting electrons to follow a direction aligned with negative tracks and rotated with respect to positive tracks~\cite{Acharya:2017lco}. As a consequence, the performance of track-segment reconstruction and the likelihood calculation differs for electrons and positrons.

\subsection{Track selection}
Tracks are required to have at least 70 out of 159 clusters in the TPC and hits in at least 2 of the 6 ITS layers. A hit in the innermost layer of the SPD is specifically required to suppress background electrons from photon conversions in the detector material. Tracks are selected within the \TRD kinematic acceptance, with \pt larger than 1~\GeVc and pseudorapidity $|\eta| < 0.84$. To further suppress background electrons and tracks originating from weak decays, tracks with a Distance-of-Closest-Approach (DCA) to the primary vertex larger than 0.2~cm in the transverse plane and 0.4~cm along the beam axis are rejected. The transverse DCA resolution is better than 75 $\mu$m for charged particles with \pt $>$ 1 GeV/c~\cite{ALICE:2018fvj}. Additionally, to reduce contributions from photon conversions, any track forming an opposite-sign pair with invariant mass below 0.05~\GeVmass is excluded. This selection is verified with MC simulations to have a negligible impact on the charmonium signal.

Good quality tracks are identified as electrons (and positrons) based on their \dEdx measured in the TPC. Specifically, the variable $n_{\sigma_{i}}$ is used. It represents the deviation (in units of standard deviations) of the measured energy loss from the expected value for a given particle hypothesis $i$. Electron candidates are required to satisfy $-2 < n_{\sigma_{\mathrm e}} < 3$. To reject contamination from proton and charged pion ($\pi$), tracks with momenta $p < 5$~\GeVc and $|n_{\sigma_{\mathrm p, \pi}}| < 3$ are vetoed. For tracks with momenta $p > 5$~\GeVc, a looser criterion of $|n_{\sigma_{\mathrm p, \pi}}| < 2$ is applied to improve the reconstruction efficiency and retain a sufficient number of electron candidates.

\subsection{Signal extraction}
The raw yields of \jpsi and \psis in each \pt interval are obtained from the invariant-mass distributions of opposite-sign (OS) electron--positron pairs (\mee) formed within the same event (SE). These distributions contain contributions from the signal (\jpsi and \psis) as well as background. The dominant background consists of electron--positron pairs from uncorrelated processes, mostly semi-leptonic heavy-flavor decays combined with light vector meson decays.

This uncorrelated combinatorial background is estimated using the mixed-event (ME) technique, as employed in previous measurements~\cite{ALICE:2021dtt,ALICE:2018mml}.
To construct the ME background, each event is assigned to a mixing pool defined by a similar number of SPD tracklets (short track segments reconstructed in the two layers of the SPD) and a similar primary vertex position ($z$-vertex). Mixing is performed using events within the same run, corresponding to a stable data-taking period. The OS distribution from mixed events is normalised using the ratio of the same-event like-sign to mixed-event like-sign distributions in the mass range $2 < \mee < 4$~\GeVmass, where contributions from correlated sources are expected to be small.

After subtracting the uncorrelated background using the ME technique, the remaining distribution is fitted using signal templates of \jpsi and \psis obtained from the MC simulation described in the previous section. A second-order polynomial function is used to describe the residual background mostly coming from correlated electron pairs from semi-leptonic decays of heavy-flavor hadrons~\cite{ALICE:2018gev}. The raw number of \jpsi and \psis is obtained by integrating the signal distributions in the invariant-mass intervals 2.92 $< \mee <$ 3.16 \GeVmass and 3.52 $< \mee <$ 3.76 \GeVmass, respectively. The quality of the fit, expressed as the $\chi^2/\mathrm{NDF}$, ranged from 1.21 to 1.10 across the considered \pt intervals, decreasing from low to high electron--positron pair \pt (\ptee).

\begin{figure}[htb]
    \centering
        \includegraphics[width=0.98\linewidth]{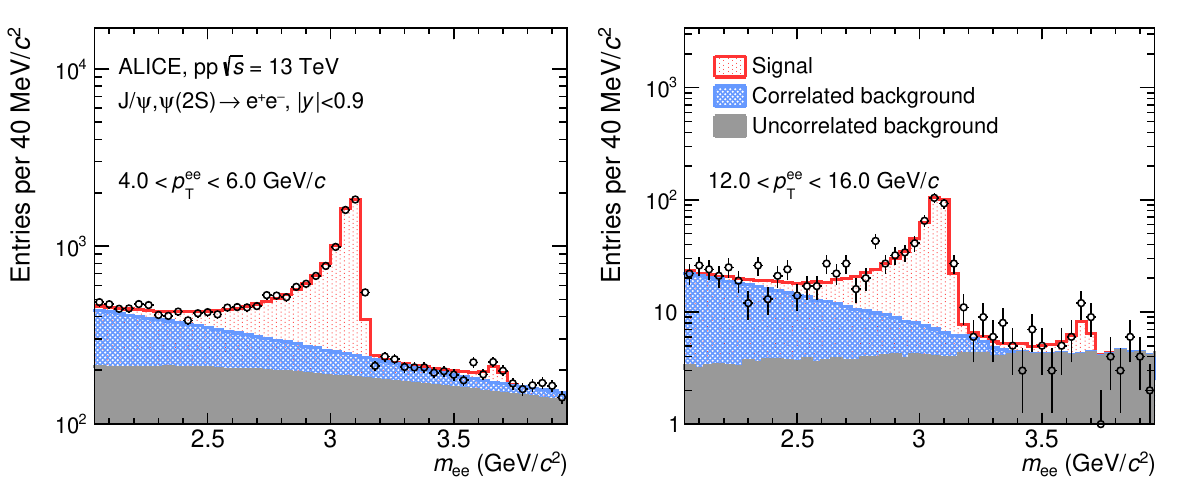}
    \caption{Invariant-mass distributions of \ee pairs in the same event for the lowest (left) and the highest (right) \ptee intervals. Superimposed to the data are the fits containing the signals of \jpsi and \psis and the correlated (blue area) and uncorrelated (gray area) background sources (see text).}
    \label{fig:invmass}
\end{figure}

Figure~\ref{fig:invmass} shows the invariant-mass distributions of electron--positron pairs for the lowest (left) and highest (right) \ptee intervals.
These distributions include contributions from signal (\jpsi and \psis), correlated background (primarily from heavy-flavor decays), and uncorrelated combinatorial background estimated with the ME technique. The uncorrelated background is shown as the gray area. Although the signal extraction fit is performed after subtracting this ME background, the ME component is added back and overlaid in the figure for illustrative purposes. The total fit function is decomposed into the correlated background (blue shaded area) and the signal peaks (red). Even in the lowest dielectron pair-\pt interval (4 $< \ptee <$ 6~\GeVc), the invariant-mass distribution displays a visible \psis peak with a statistical significance of 3.9 $\sigma$. This constitutes the first measurement of \psis at midrapidity at the LHC at such low \pt. 
The significance of the \psis signal is improved by the larger integrated luminosity available at low \pt compared to the MB data and by the improved electron selection due to the \TRD online PID. The lower \pt reach is constrained by the signal-to-background ratio rather than by the kinematic threshold imposed by the trigger requirements.

\subsection{Corrections}
\label{sec:correction}
\begin{figure}[htb]
    \centering 
        \includegraphics[width=0.98\linewidth]{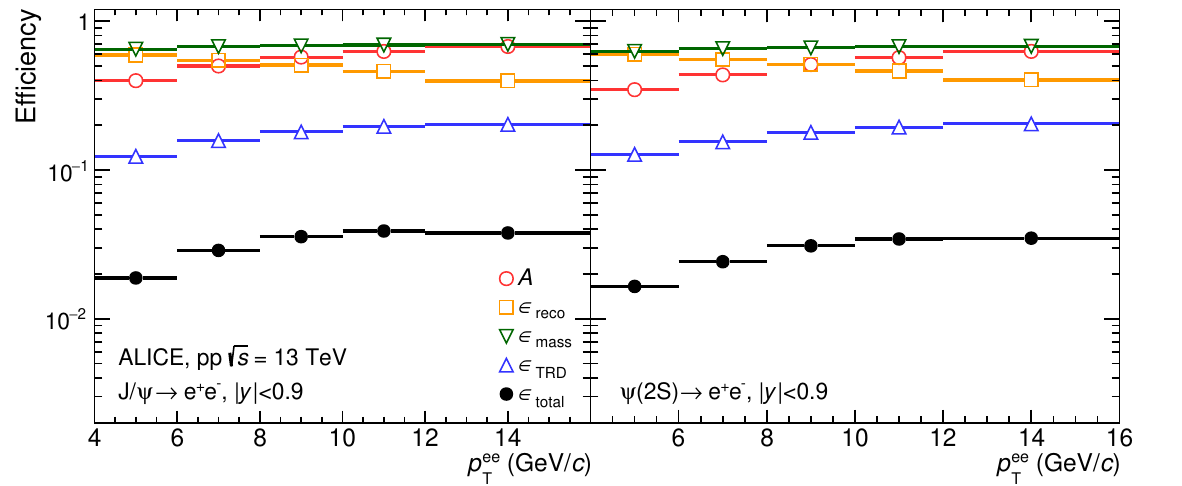}
    \caption{Acceptance and efficiency as a function of \ptee for inclusive \jpsi (left) and \psis (right). 
    In addition to the total efficiency, its individual contributions are shown.}
    \label{fig:eff}
\end{figure}

The corrections for acceptance and efficiency ($\it{A} \times \epsilon$) are obtained from MC simulations and include the product of the following components: (i) the kinematic acceptance ($\it{A}$); (ii) the reconstruction efficiency ($\epsilon_{\mathrm{reco}}$), which accounts for both tracking and PID efficiencies; (iii) the efficiency due to the finite integration mass window used for extracting the \jpsi and \psis yields ($\epsilon_{\mathrm{mass}}$); and (iv) the \TRD-trigger efficiency ($\epsilon_{\mathrm{TRDtrg}}$), which represents the probability that an event containing a charmonium decay satisfies the \TRD-trigger condition.
The acceptance and total efficiency, with its individual contributions, are shown in Fig.~\ref{fig:eff} for \jpsi (left) and \psis (right).

The \TRD-trigger efficiency of \jpsi and \psis is emulated in MC simulations by implementing the same online trigger algorithm as described above. To evaluate the \TRD-trigger efficiency of charmonia, only events containing prompt \jpsi and \psis are used to avoid an artificial enhancement of the trigger probability from additional electrons produced in beauty-hadron decays. In addition, the efficiency is also evaluated by selecting only those candidates for which the trigger is fired by one of their own decay electrons. In the simulations, both charmonia are embedded in the same event and forced to decay into dielectrons, which increases the number of lepton pairs compared to real data and consequently raises the likelihood that at least one track fulfills the \TRD-trigger condition. This contribution is found to be independent of \pt. The underlying-event contribution included in $\epsilon_{\mathrm{TRDtrg}}$ is therefore also taken into account in the charmonium trigger efficiency.

\subsection{Systematic uncertainties}
Table~\ref{tableUncAll} summarizes the sources of systematic uncertainty for the \pt-differential cross section measurements of \jpsi, \psis, and the \psis-to-\jpsi ratio. In order to obtain stable estimates and minimize the impact of statistical fluctuations, the uncertainties are evaluated in two broader \pt intervals: 4--8 and 8--16~\GeVc. These intervals are selected based on the similarity of background shapes and signal-to-background ratios (S/B). The uncertainties are evaluated by varying the specific criteria relevant to each source and calculating the root mean square of the resulting distributions. The total systematic uncertainty used for the cross section is calculated as the quadratic sum of all individual sources, excluding contributions from the branching ratios and integrated luminosity.

\begin{table}[ht]
    \centering
    \caption{Summary of systematic uncertainties for the \jpsi and \psis yields, and the \psis-to-\jpsi yield ratio, evaluated in the \pt intervals 4--8 and 8--16~\GeVc.}
	\begin{tabular}{c|cc|cc|cc}
              & \multicolumn{2}{c|} {\jpsi} & \multicolumn{2}{c|} {\psis} & \multicolumn{2}{c} {\psis/\jpsi}\\
            \hline
              $p_{\rm T}$ (GeV/$c$) & 4--8 & 8--16 & 4--8& 8--16& 4--8& 8--16\\
             \hline
              Tracking &  {5.4\%} &  {7.1\%} &  {5.4\%} &  {7.1\%} & \multicolumn{2}{c} {cancels}\\
              PID &  {1.2\%} &  {3.4\%} &  {1.2\%} &  {3.4\%} & \multicolumn{2}{c} {cancels}\\
              Signal extraction &  {1.9\%} &  {1.9\%} &  {3.3\%} &  {2.9\%} &  {3.8\%} & {4.4\%}\\
              Fit &  {1.6\%} &  {1.8\%} &  {9.9\%} &  {9.2\%} &  {10.0\%} & {9.4\%}\\
              MC $p_{\rm T}$ shape & \multicolumn{2}{c|} {2.8\%}  &  \multicolumn{2}{c|} {2.0\%}  & \multicolumn{2}{c} {3.4\%}\\
              \TRD-trigger efficiency & \multicolumn{2}{c|} {5.0\%} & \multicolumn{2}{c|} {5.0\%} & \multicolumn{2}{c} {cancels}\\
            \hline
            \hline
              Branching ratio & \multicolumn{2}{c|} {0.54\%\cite{pdg}} & \multicolumn{2}{c|} {2.8\%\cite{pdg}} & \multicolumn{2}{c} {2.9\%}\\
              Luminosity & \multicolumn{4}{c|} {1.8\%\cite{ALICE:2021leo}} &  \multicolumn{2}{c} {cancels}\\
            \end{tabular}
    \label{tableUncAll}
\end{table}

The systematic uncertainty of the tracking efficiency is determined as the quadratic sum of the uncertainties from the ITS--TPC matching efficiency and from the efficiency of the track-quality selections. The uncertainty of the ITS--TPC matching efficiency is derived from residual differences between data and MC simulations for single-track matching, as described in Ref.~\cite{ALICE:2017olh}. Since this uncertainty is found to be independent of the particle species, it is propagated to dielectron pairs from \jpsi and \psis decays, yielding values between 4\% and 5\% depending on \pt. The uncertainty of the track quality is obtained by varying the corresponding selection criteria. 

The systematic uncertainty on PID is determined by varying the PID selection criteria. Both the electron-identification cuts and the hadron-rejection criteria are varied in data and MC simulations.

For \psis, the same tracking and PID systematic uncertainties as for \jpsi are assigned. This approach reduces the impact of statistical fluctuations in the uncertainty estimation. To validate this assumption, the reconstruction efficiencies (including tracking and PID) for electrons from \jpsi and \psis decays are compared. The ratio of the efficiencies is found to be constant for \pt $>$ 2~\GeVc, indicating similar detector response and selection efficiency in this \pt region. Therefore, assigning the same tracking and PID systematic uncertainties to both charmonium states is justified. Consequently, these uncertainties cancel in the \psis-to-\jpsi ratio.

The systematic uncertainty related to the signal extraction is determined by varying the limits of the invariant-mass interval of the \ee pair for signal integration. This is done by extending or reducing the integration interval by 40 \MeVmass to 80 \MeVmass in upper and lower direction. The data are corrected for the fraction of signal outside of the integration interval.

The systematic uncertainty related to the invariant-mass fit is determined by varying the functions corresponding to the correlated background and signal shape and the fit range. Even in the high-\pt region, the S/B for \psis is around 2, indicating that the background contribution remains non-negligible, and the background fit still contributes significantly to the overall uncertainty.

The initial \pt distributions of \jpsi and \psis used in the MC simulations are reweighted to match measurements of the \jpsi at midrapidity and of the \psis at forward rapidity~\cite{ALICE:2021dtt,ALICE:2017leg}. The systematic uncertainty associated with the input \pt distributions is evaluated by varying these spectra and propagating the effect to the acceptance and efficiency corrections. As alternative input shapes, the \jpsi \pt distribution measured at forward rapidity~\cite{ALICE:2017leg} is used for \jpsi, while the midrapidity \jpsi distribution scaled according to the $m_{\mathrm{T}}$-scaling prescription~\cite{ALICE:2017leg,Gatoff:1992cv} is used for \psis.

The systematic uncertainty on the \TRD-trigger efficiency for \jpsi and \psis is obtained by varying the likelihood threshold used in the trigger decision within the integer granularity of the front-end electronics, following the procedure described in Ref.~\cite{ALICE:2022zig}. This uncertainty accounts for possible residual differences between data and simulation in the \TRD PID response due to imperfect modelling of detector conditions, such as pressure and gas composition. In addition, in the determination of the \TRD-trigger efficiency for electrons (Eq.~\ref{eq:slEff}), both correlated and uncorrelated electron contributions are included. In dielectron decays from charmonia, the trigger decision can be fired by either the electron or the positron in the pair. Consequently, an electron that does not directly satisfy the trigger requirement may still appear in a triggered event due to its correlated partner. This effect leads to a slight \pt-dependent enhancement of the estimated efficiency, increasing the absolute trigger efficiency by up to about one percentage point in the simulation. Since correlated and uncorrelated contributions cannot be disentangled in real data, this variation is introduced to conservatively cover the associated uncertainty and any residual differences between data and simulation.

The uncertainty of the integrated luminosity includes contributions from the beam luminosity estimated via a van der Meer scan (1.8\%)~\cite{ALICE:2021leo} and from the \TRD trigger normalisation factor (0.2\%).
These two components are added in quadrature and cancel in the \psis-to-\jpsi ratio. The uncertainty of the branching ratios to the dielectron channels is taken from Ref.~\cite{pdg}.

\section{Results}\label{results}

\subsection{\texorpdfstring{$\pt$}{pt}-differential production cross section}

\begin{figure}[htb]
    \centering
        \includegraphics[width=0.7\linewidth]{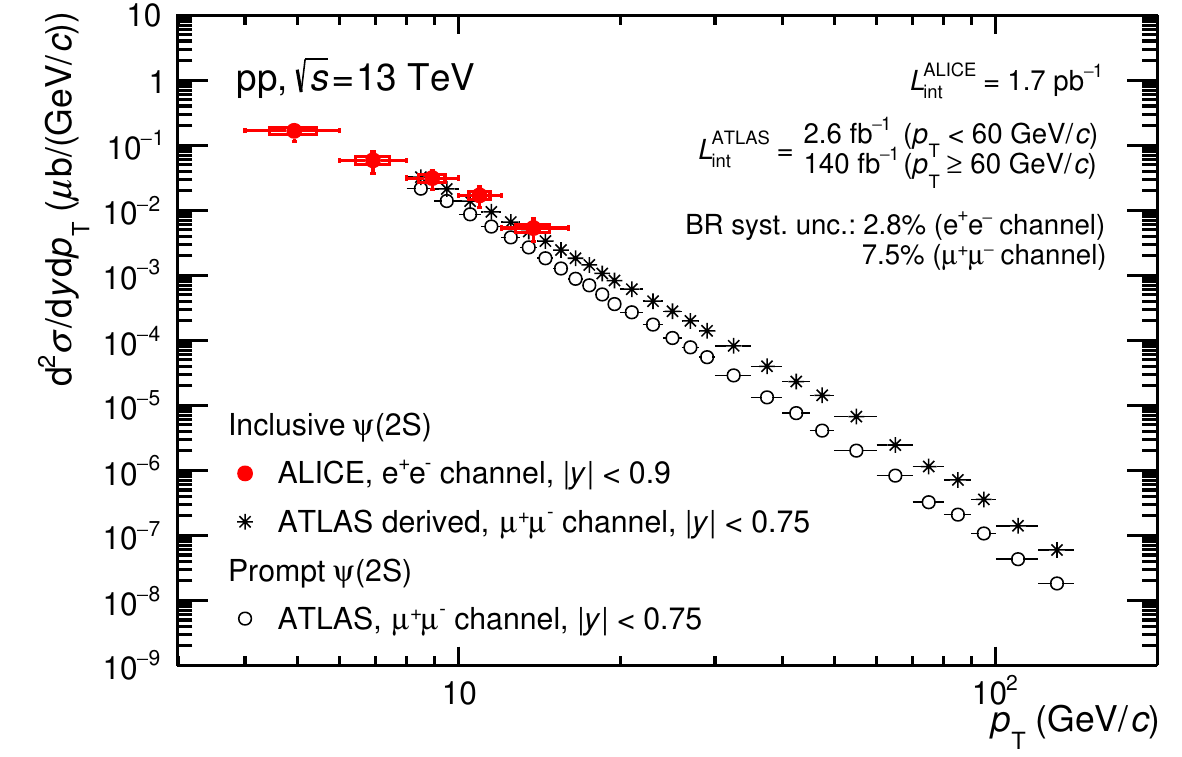}
    \caption{
    \pt-differential inclusive \psis production cross section at midrapidity in \pp collisions at \thirteen. Statistical and systematic uncertainties are shown as error bars and boxes, respectively, and data points are plotted according to the Lafferty–Wyatt prescription~\cite{Lafferty:1994cj}. The data points for the inclusive production are obtained by summing the published prompt and non-prompt measurements~\cite{ATLAS:2023qnh}. The statistical and systematic uncertainties are combined in quadrature. Uncertainties related to integrated luminosity are not shown.
    }
    \label{fig:XsectionComp}
\end{figure}

In Fig.~\ref{fig:XsectionComp}, the \pt-differential inclusive \psis production cross section measured by ALICE is compared with results from the ATLAS Collaboration~\cite{ATLAS:2023qnh}, covering \pt ranges of 8 $<$ \pt $<$ 140~\GeVc ($|y| < 0.75$). For the ALICE data, statistical and systematic uncertainties are represented by error bars and boxes, respectively.
The uncertainties on branching ratios of \psis to electrons and muons of about 2.8\% and 7.5\%, respectively, are not included in the boxes. In order to account for the non-linearity of the steeply falling spectrum within each \pt interval, the ALICE data points in the figure are placed at the interval-averaged \pt value following the Lafferty-Wyatt prescription~\cite{Lafferty:1994cj}. To enable a direct comparison to the inclusive ALICE measurements, the ATLAS data points are obtained by summing the prompt and non-prompt cross sections reported in Ref.~\cite{ATLAS:2023qnh}. In this procedure, the uncertainties of the prompt and non-prompt components are assumed to be correlated when forming the inclusive uncertainty. The statistical and systematic uncertainties of the resulting inclusive points are then added in quadrature and shown as error bars for visual clarity. The integrated luminosity uncertainty is now shown in the figure; it amounts to 1.8\% for ALICE and to 1.13\% (\pt $<$ 60~\GeVc) and 0.83\% (\pt $\geq$ 60~\GeVc) for ATLAS~\cite{ATLAS:2023qnh}.

In the overlapping \pt region, the inclusive \psis production cross section is fully consistent with the derived ATLAS data in the range 8--16~\GeVc within the uncertainties. The ALICE measurement extends the accessible \pt range from 8 down to 4~\GeVc. 

\begin{figure}[htb]
    \centering
    \subfigure{
        \includegraphics[width=0.48\linewidth]{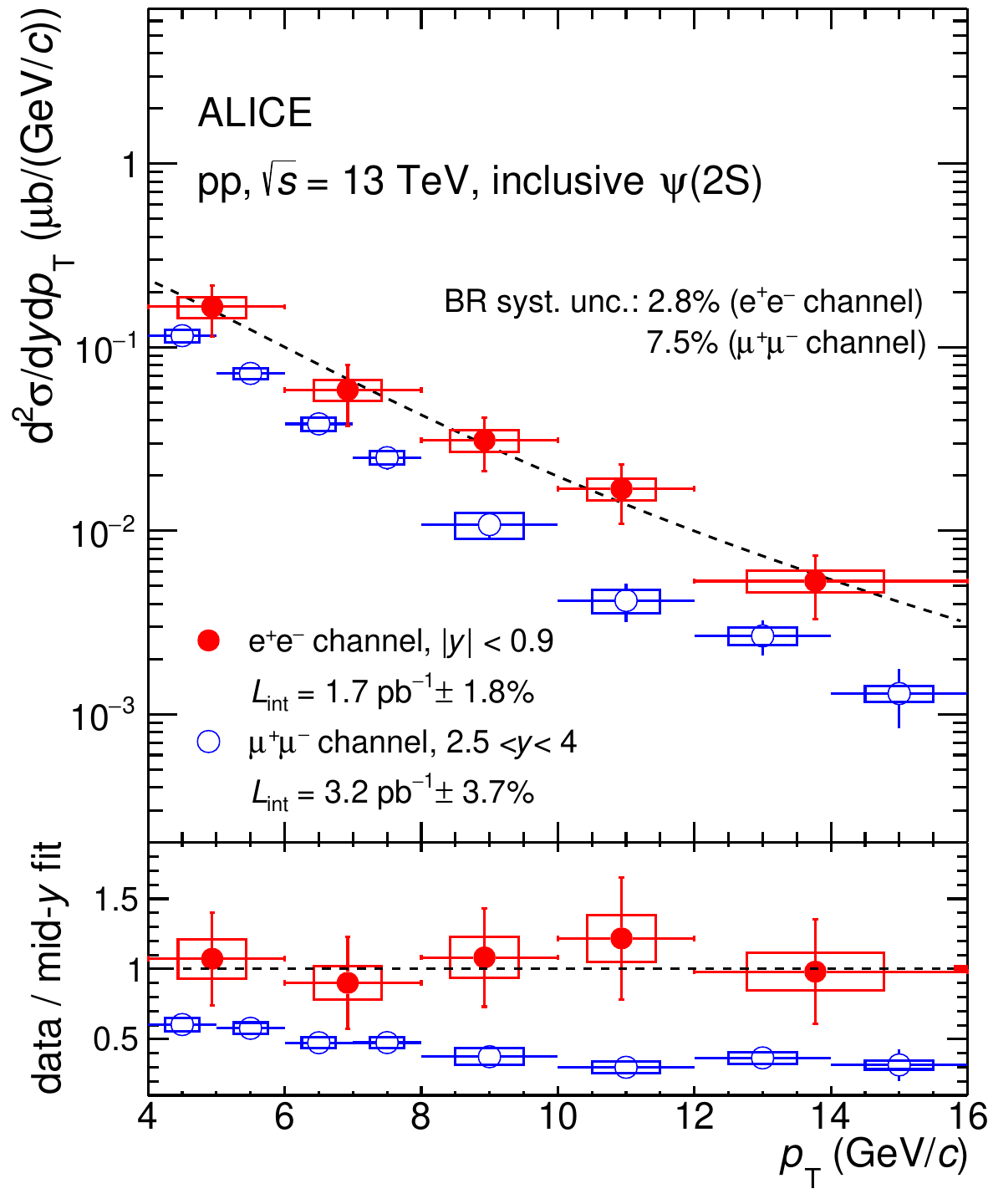}
    }
    \subfigure{
        \includegraphics[width=0.48\linewidth]{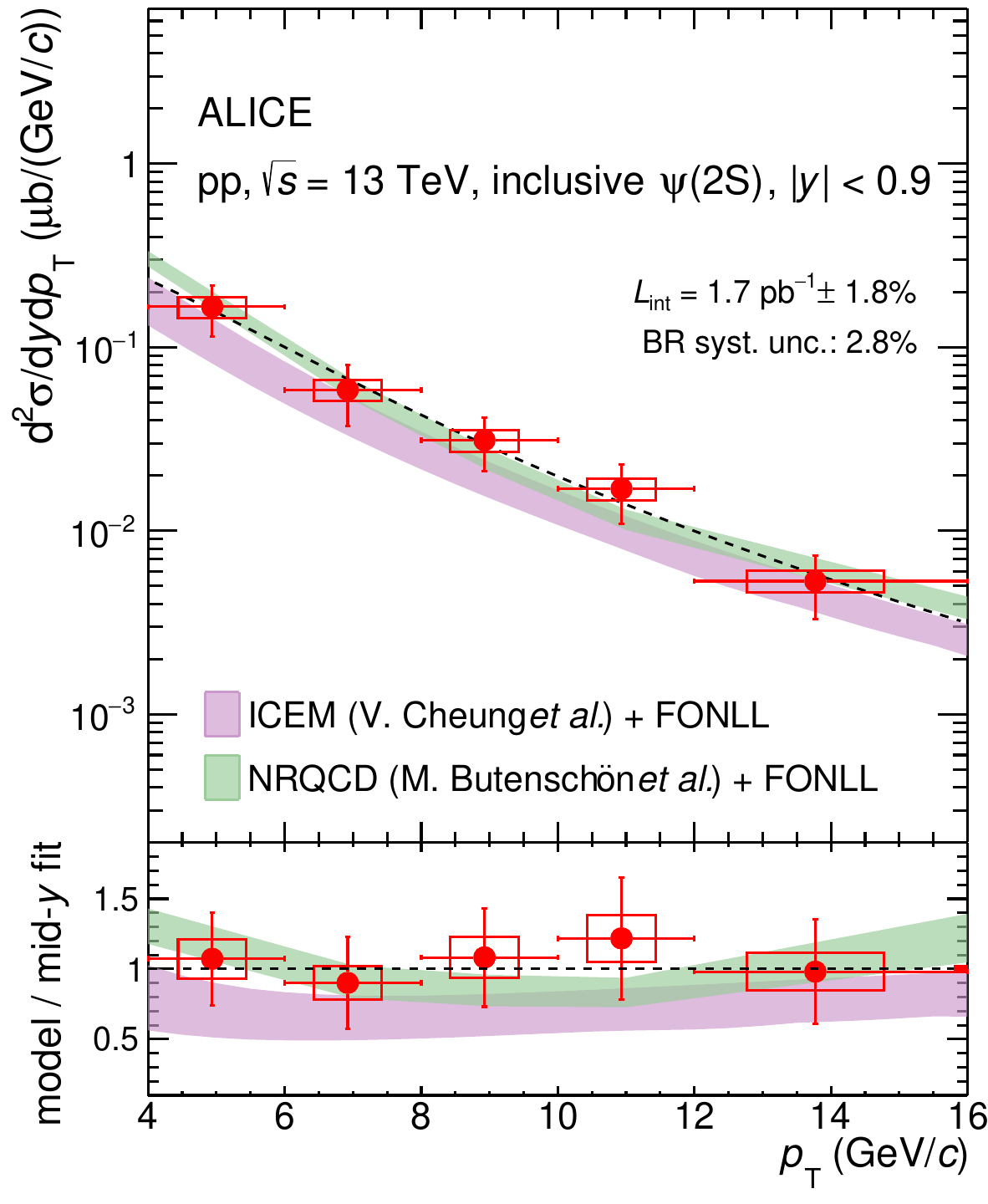}
    }
    \caption{\pt-differential inclusive \psis production cross section at midrapidity in \pp collisions at \thirteen is compared with ALICE results at forward rapidity~\cite{ALICE:2017leg} (left) and with the model calculations from ICEM~\cite{icem} and NRQCD~\cite{nrqcd} (right). Both calculations include a non-prompt contribution from beauty-hadron decays via FONLL~\cite{fonll}. Data points are plotted according to the Lafferty–Wyatt prescription~\cite{Lafferty:1994cj}. The lower panels show the ratios of the data points or model calculations to the power-law fit to the midrapidity data.}
    \label{fig:xsection}
\end{figure}

In Fig.~\ref{fig:xsection}, the inclusive \pt-differential \psis production cross section in \pp collisions at \thirteen at midrapidity ($|y| < 0.9$) is shown and fitted with a power-law function
\begin{equation}
    f(\pt) = K \times \frac{\pt}{{(1+(\pt/p_{0})^{2})}^{n}}\, ,
\end{equation}
with $K$, $p_{0}$, and $n$ as free parameters. The fit describes the data well, with $\chi^{2}/\mathrm{NDF} = 0.45$.

The comparison of \jpsi production between mid and forward rapidity was studied in Ref.~\cite{ALICE:2021dtt} and showed a hardening of the \pt spectrum at midrapidity compared to forward rapidity. The corresponding comparison for \psis is presented in this paper. In the left panel of Fig.~\ref{fig:xsection}, the \psis production cross section at midrapidity is compared to the measurement at forward rapidity~\cite{ALICE:2017leg}. 
The lower panel shows the ratios of the \psis measurements at midrapidity and forward rapidity to the power-law fit to the midrapidity data. 
The uncertainty boxes in the lower panel represent the quadratic sum of the systematic uncertainties of the data points and the uncertainty of the fit-function mean in the given \pt interval. The ratio of the measurements at forward rapidity to the fit to the midrapidity results decreases by about a factor of two between 4.5 and 15~\GeVc, indicating a harder \psis spectrum at midrapidity compared to forward rapidity. This is consistent with the behavior observed for the \jpsi spectra~\cite{ALICE:2021dtt}.

In the right panel of Fig.~\ref{fig:xsection}, the \pt-differential inclusive \psis production cross section is compared with results from two model calculations, the  ICEM~\cite{icem} model and NRQCD~\cite{nrqcd}. Both calculations only include prompt \psis production; therefore, a non-prompt \psis component due to weak decays of b hadrons is added to these model calculations, obtained by using the Fixed-Order Next-to-Leading-Logarithms (FONLL) approach by Cacciari et al. from Ref.~\cite{fonll}. The non-prompt fraction, defined as the ratio of non-prompt to inclusive \psis production, increases from 10\% at low \pt to 49\% at \pt = 16~\GeVc.

The lower panel of the right-hand side of the figure shows again the ratios to the power-law fit to the \psis data. The bands represent the uncertainties of the model calculations. These uncertainties mostly arise from variations in the renormalisation and factorization scales and the charm-quark mass. The NLO NRQCD calculation by Butenschoen et al.~in Ref.~\cite{nrqcd} employs the factorization formalism, where the production cross section is decomposed into short-distance coefficients, calculated perturbatively, and LDMEs. These are fitted to experimental data and account for contributions from various intermediate Fock states. They are assumed to be universal. In contrast, the ICEM by Cheung et al. from Ref.~\cite{icem} uses the $k_{\mathrm{T}}$-factorization approach and includes a single normalisation parameter to account for long-distance effects~\cite{Nelson:2012bc} instead of using LDMEs. 

The NRQCD results have a small uncertainty and are in agreement with the experimental data over the full \pt range, similar to observations made already at forward rapidity~\cite{ALICE:2017leg}. The ICEM predictions are also consistent with the data within uncertainties, although the calculation tends to lie near the lower edge of the measured cross sections. This behaviour is mainly due to the harder \pt spectrum predicted by the model and the fitting procedure used to determine the overall normalisation, which is strongly influenced by low-\pt CDF data~\cite{CDF:2009kwm}.

\subsection{\texorpdfstring{$\ensuremath{\mathrm{\psi(2S)}}$}{psis}-to-\texorpdfstring{$\ensuremath{\mathrm{J/\psi}}$}{jpsi} cross section ratio}
\begin{figure}[htb]
    \centering
        \includegraphics[width=0.7\linewidth]{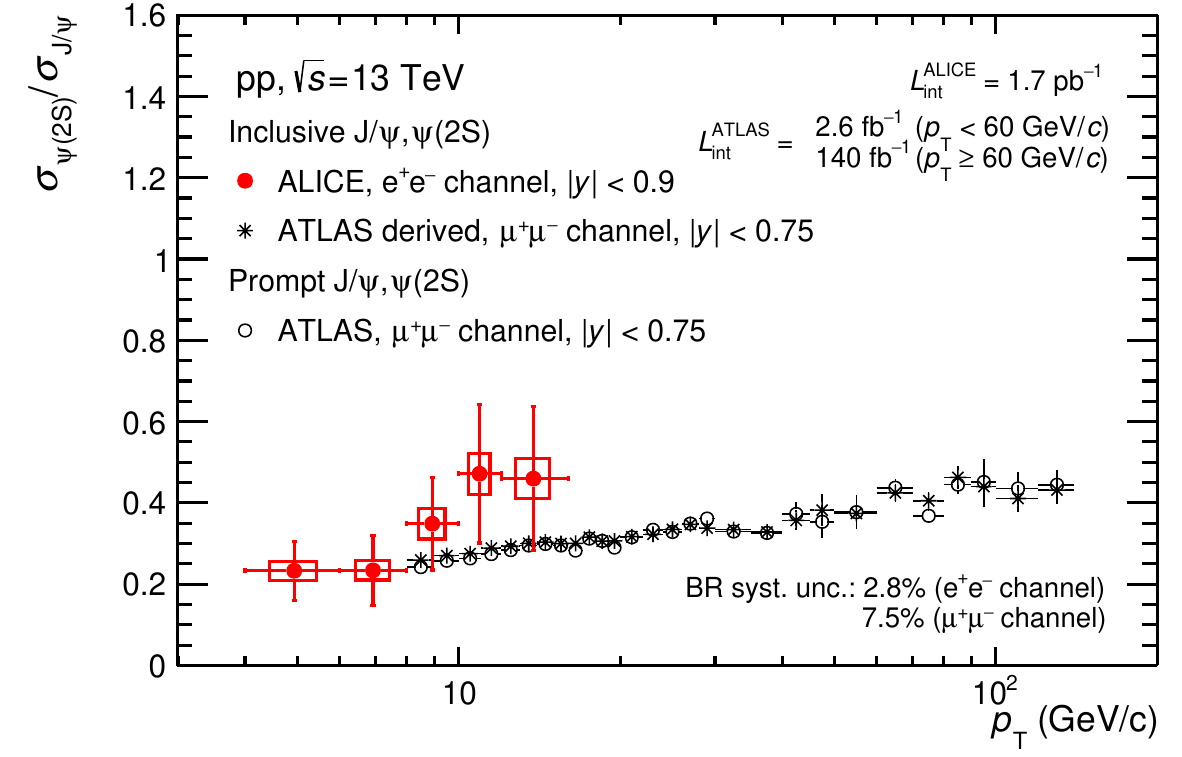}
    \caption{\pt-differential \psis-to-\jpsi production cross section ratio measured by the ALICE and the ATLAS Collaborations at midrapidity in \pp collisions at \thirteen~\cite{ATLAS:2023qnh}. For the ALICE Collaboration, statistical and systematic uncertainties are shown as error bars and boxes, respectively, and data points are plotted according to the Lafferty–Wyatt prescription~\cite{Lafferty:1994cj}.
    Inclusive \psis-to-\jpsi points from the ATLAS Collaboration (solid black circles) are obtained by summing the published prompt and non-prompt components of each state~\cite{ATLAS:2023qnh}, and taking their ratio. Statistical and systematic uncertainties are treated as uncorrelated and combined in quadrature.
    }
    \label{fig:ratioComp}
\end{figure}

Figure~\ref{fig:ratioComp} shows the inclusive \psis-to-\jpsi cross section ratio measured by the ALICE Collaboration in \pp collisions at \thirteen, compared with results from ATLAS~\cite{ATLAS:2023qnh}. The \jpsi cross section used in the ratio is obtained from the present analysis and is verified to be fully consistent with previous ALICE measurements~\cite{ALICE:2017leg}.
The statistical and systematic uncertainties of the ALICE results are represented by error bars and boxes, respectively. The luminosity uncertainty cancels out in the ratio, and the uncertainties due to the branching ratios are shown in the legend.
Inclusive \psis-to-\jpsi data (solid black circles) from the ATLAS Collaboration are reconstructed by first summing the published prompt and non-prompt components of each state to obtain their inclusive cross sections~\cite{ATLAS:2023qnh}, and then taking their ratio. The statistical and systematic uncertainties of the two components are treated as uncorrelated and combined in quadrature. The inclusive \psis-to-\jpsi production cross section ratio measured by ALICE is consistent with the corresponding result from ATLAS~\cite{ATLAS:2023qnh}.

\begin{figure}[htb]
    \centering
    \subfigure{
        \includegraphics[width=0.485\linewidth]{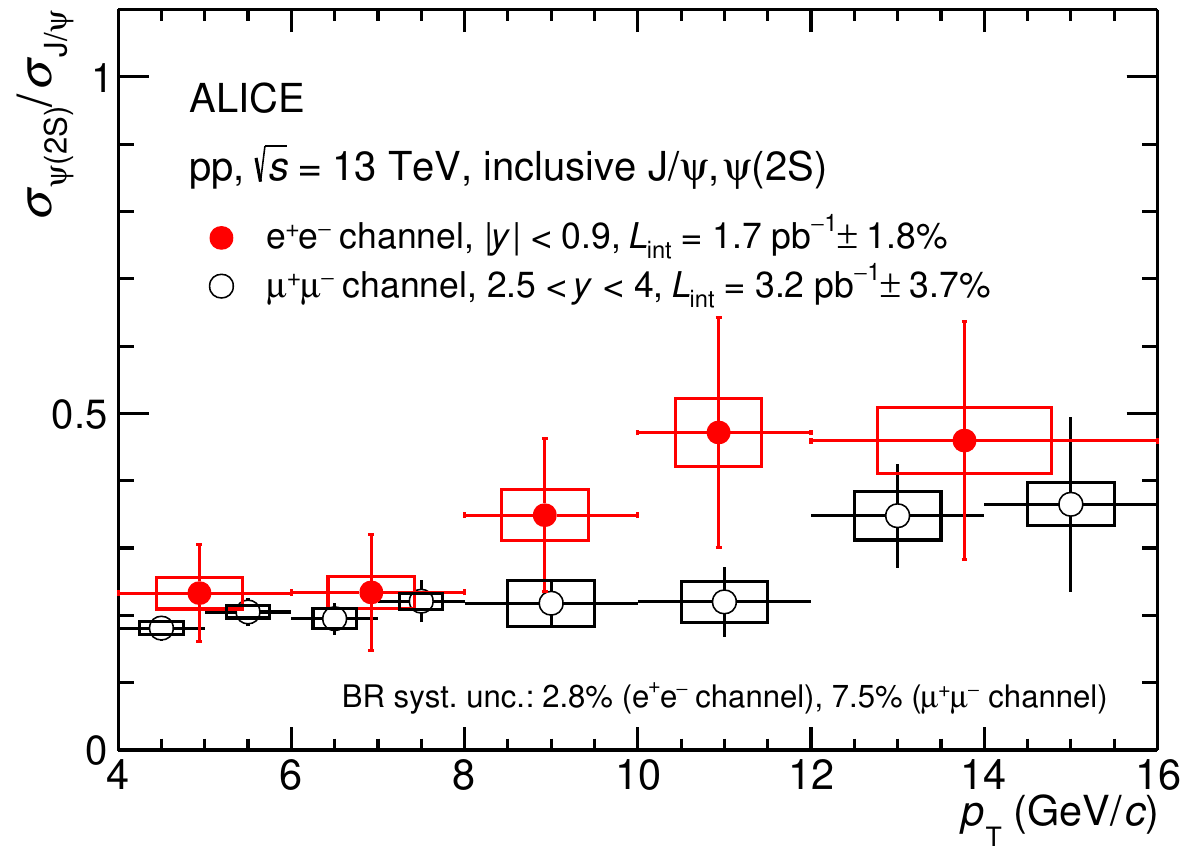}
    }
    \subfigure{
        \includegraphics[width=0.485\linewidth]{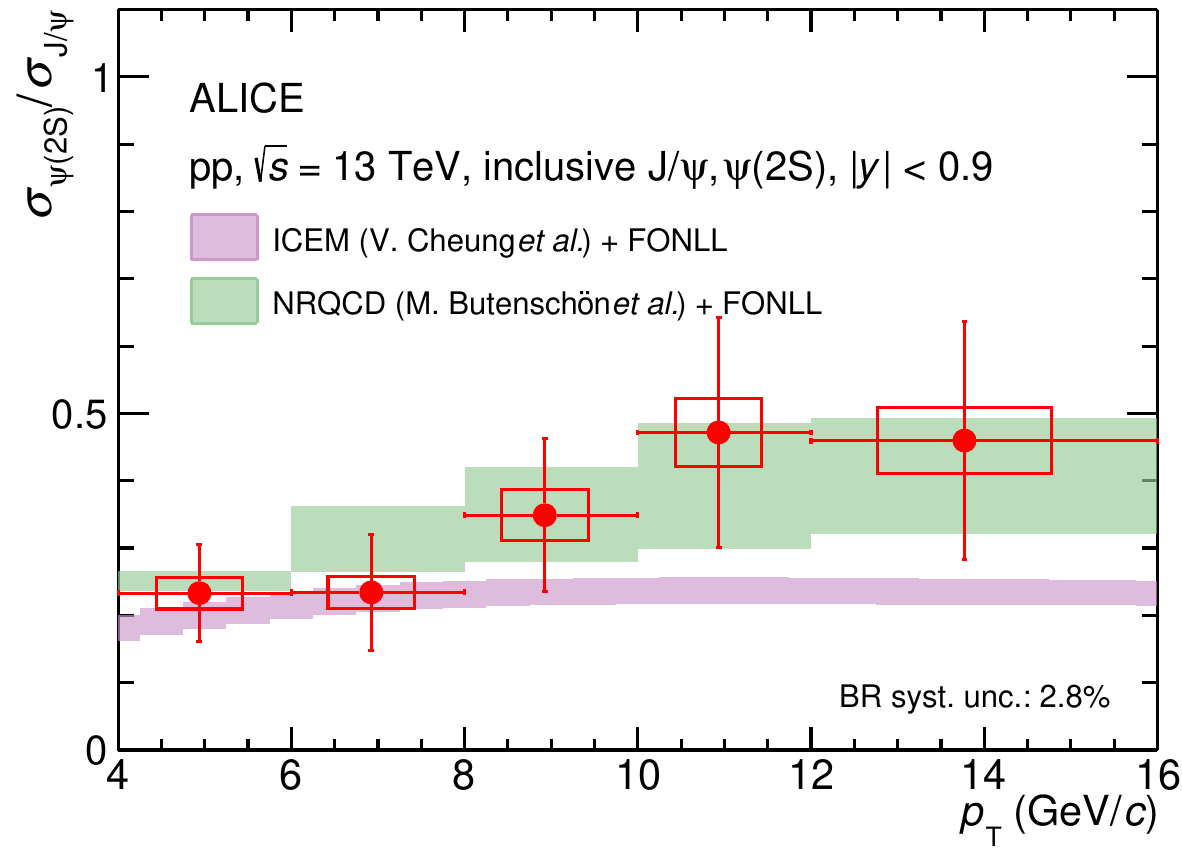}
    }
    \caption{\pt-differential \psis-to-\jpsi production cross section ratio at midrapidity in \pp collisions at \thirteen compared with a measurement at forward rapidity in \pp collisions at the same collision energy~\cite{ALICE:2017leg} (left), and compared with the model calculations~\cite{icem, nrqcd, fonll} (right).
    The error bars indicate the statistical uncertainties, and the boxes indicate the systematic uncertainties. The data points are plotted at the effective \pt values evaluated according to the Lafferty-Wyatt prescription~\cite{Lafferty:1994cj}.
    The widths of the bands reflect the theoretical uncertainties.
    } 
    \label{fig:ratio}
\end{figure}

In the left panel of Fig.~\ref{fig:ratio}, the \pt-differential \psis-to-\jpsi cross section ratio is compared with measurements at forward rapidity in \pp collisions at the same center-of-mass energy~\cite{ALICE:2017leg}. The two measurements are compatible within current uncertainties, indicating no significant rapidity dependence in the \psis-to-\jpsi cross section ratio. A mild increase of the inclusive \psis-to-\jpsi cross section ratio as a function of \pt is observed in both cases, although the uncertainties are large, especially at high \pt. The observed trend may be partially due to feed-down contributions from higher mass states such as \psis and $\chi_{\mathrm{c}}$.

In the right panel of Fig.~\ref{fig:ratio}, the inclusive \psis-to-\jpsi cross section ratio is compared with model calculations. The uncertainties of theoretical calculations from variations of scales and masses cancel out in the ratio. Theory bands are obtained by taking the ratio of the upper and lower bounds of \psis and \jpsi model calculations.  
In models like ICEM, \jpsi from feed-down from higher charmonium states has softer \pt spectra than their parent particles, leading to a slightly rising ratio with increasing \pt.
In contrast, NRQCD predicts genuinely harder \pt spectra for heavier states, reflecting hard scattering kinematics. The observed trend may result from a combination of small higher-state feed-down effects and intrinsic production dynamics.  
The NRQCD model shows good agreement within the relatively large model uncertainties over the entire measured \pt range, consistent with observations at forward rapidity~\cite{ALICE:2017leg}. The size of the uncertainty of the NRQCD calculation on the \psis-to-\jpsi cross section ratio arises from the LDMEs fits, which start at low \pt and include a large number of datasets. The ICEM calculation describes the \jpsi cross section within uncertainties in the $4 < \pt < 16$~\GeVc range~\cite{ALICE:2017leg}, while underestimating the \psis yield. Since ICEM assumes a \pt dependence of prompt charmonium production independent of collision energy and rapidity, it inherently predicts an almost flat \psis-to-\jpsi ratio, whereas the data show an increasing trend with \pt.

\section{Summary}\label{summary}

The inclusive \pt-differential production cross section of \psis mesons and the ratio to the \jpsi cross section have been measured at midrapidity ($|y| <$ 0.9) in pp collisions at \s = 13 TeV for the first time with the ALICE detector at the LHC. The measurements are reported for the \pt range of 4 to 16~\GeVc using data collected with the online single-electron trigger from the Transition Radiation Detector. This analysis extends the accessible rapidity and \pt ranges compared to previous studies at the LHC, providing new insights into charmonium production at relatively low \pt. Comparison with theoretical predictions shows that NRQCD calculations describe the data quantitatively over the entire measured \pt range.
A mild rise in the \psis-to-\jpsi cross section ratio with increasing \pt is observed, consistent within uncertainties with the trend predicted by NRQCD. While the ICEM results are also consistent within errors with the measured \psis spectrum, 
the ratio of \psis-to-\jpsi shows a significantly weaker increase with \pt, which is inherent to the ICEM approach.
The ALICE results of the \psis cross section and the \psis-to-\jpsi cross section ratio are consistent with the inclusive measurements derived from ATLAS results in the overlapping \pt region and extend the coverage from 8~\GeVc down to 4~\GeVc, a region not accessible in other experiments.

\newenvironment{acknowledgement}{\relax}{\relax}
\begin{acknowledgement}
\section*{Acknowledgements}

The ALICE Collaboration would like to thank all its engineers and technicians for their invaluable contributions to the construction of the experiment and the CERN accelerator teams for the outstanding performance of the LHC complex.
The ALICE Collaboration gratefully acknowledges the resources and support provided by all Grid centres and the Worldwide LHC Computing Grid (WLCG) collaboration.
The ALICE Collaboration acknowledges the following funding agencies for their support in building and running the ALICE detector:
A. I. Alikhanyan National Science Laboratory (Yerevan Physics Institute) Foundation (ANSL), State Committee of Science and World Federation of Scientists (WFS), Armenia;
Austrian Academy of Sciences, Austrian Science Fund (FWF): [M 2467-N36] and Nationalstiftung f\"{u}r Forschung, Technologie und Entwicklung, Austria;
Ministry of Communications and High Technologies, National Nuclear Research Center, Azerbaijan;
Rede Nacional de Física de Altas Energias (Renafae), Financiadora de Estudos e Projetos (Finep), Funda\c{c}\~{a}o de Amparo \`{a} Pesquisa do Estado de S\~{a}o Paulo (FAPESP) and The Sao Paulo Research Foundation  (FAPESP), Brazil;
Bulgarian Ministry of Education and Science, within the National Roadmap for Research Infrastructures 2020-2027 (object CERN), Bulgaria;
Ministry of Education of China (MOEC) , Ministry of Science \& Technology of China (MSTC) and National Natural Science Foundation of China (NSFC), China;
Ministry of Science and Education and Croatian Science Foundation, Croatia;
Centro de Aplicaciones Tecnol\'{o}gicas y Desarrollo Nuclear (CEADEN), Cubaenerg\'{\i}a, Cuba;
Ministry of Education, Youth and Sports of the Czech Republic, Czech Republic;
The Danish Council for Independent Research | Natural Sciences, the VILLUM FONDEN and Danish National Research Foundation (DNRF), Denmark;
Helsinki Institute of Physics (HIP), Finland;
Commissariat \`{a} l'Energie Atomique (CEA) and Institut National de Physique Nucl\'{e}aire et de Physique des Particules (IN2P3) and Centre National de la Recherche Scientifique (CNRS), France;
Bundesministerium f\"{u}r Forschung, Technologie und Raumfahrt (BMFTR) and GSI Helmholtzzentrum f\"{u}r Schwerionenforschung GmbH, Germany;
National Research, Development and Innovation Office, Hungary;
Department of Atomic Energy Government of India (DAE), Department of Science and Technology, Government of India (DST), University Grants Commission, Government of India (UGC) and Council of Scientific and Industrial Research (CSIR), India;
National Research and Innovation Agency - BRIN, Indonesia;
Istituto Nazionale di Fisica Nucleare (INFN), Italy;
Japanese Ministry of Education, Culture, Sports, Science and Technology (MEXT) and Japan Society for the Promotion of Science (JSPS) KAKENHI, Japan;
Consejo Nacional de Ciencia (CONACYT) y Tecnolog\'{i}a, through Fondo de Cooperaci\'{o}n Internacional en Ciencia y Tecnolog\'{i}a (FONCICYT) and Direcci\'{o}n General de Asuntos del Personal Academico (DGAPA), Mexico;
Nederlandse Organisatie voor Wetenschappelijk Onderzoek (NWO), Netherlands;
The Research Council of Norway, Norway;
Pontificia Universidad Cat\'{o}lica del Per\'{u}, Peru;
Ministry of Science and Higher Education, National Science Centre and WUT ID-UB, Poland;
Korea Institute of Science and Technology Information and National Research Foundation of Korea (NRF), Republic of Korea;
Ministry of Education and Scientific Research, Institute of Atomic Physics, Ministry of Research and Innovation and Institute of Atomic Physics and Universitatea Nationala de Stiinta si Tehnologie Politehnica Bucuresti, Romania;
Ministerstvo skolstva, vyskumu, vyvoja a mladeze SR, Slovakia;
National Research Foundation of South Africa, South Africa;
Swedish Research Council (VR) and Knut \& Alice Wallenberg Foundation (KAW), Sweden;
European Organization for Nuclear Research, Switzerland;
Suranaree University of Technology (SUT), National Science and Technology Development Agency (NSTDA) and National Science, Research and Innovation Fund (NSRF via PMU-B B05F650021), Thailand;
Turkish Energy, Nuclear and Mineral Research Agency (TENMAK), Turkey;
National Academy of  Sciences of Ukraine, Ukraine;
Science and Technology Facilities Council (STFC), United Kingdom;
National Science Foundation of the United States of America (NSF) and United States Department of Energy, Office of Nuclear Physics (DOE NP), United States of America.
In addition, individual groups or members have received support from:
FORTE project, reg.\ no.\ CZ.02.01.01/00/22\_008/0004632, Czech Republic, co-funded by the European Union, Czech Republic;
European Research Council (grant no. 950692), European Union;
Deutsche Forschungs Gemeinschaft (DFG, German Research Foundation) ``Neutrinos and Dark Matter in Astro- and Particle Physics'' (grant no. SFB 1258), Germany;
FAIR - Future Artificial Intelligence Research, funded by the NextGenerationEU program (Italy).

\end{acknowledgement}

\bibliographystyle{utphys}   
\bibliography{bibliography}

\newpage
\appendix

%
%

\section{The ALICE Collaboration}
\label{app:collab}
\begin{flushleft} 
\small

D.A.H.~Abdallah\,\orcidlink{0000-0003-4768-2718}\,$^{\rm 134}$, 
I.J.~Abualrob\,\orcidlink{0009-0005-3519-5631}\,$^{\rm 112}$, 
S.~Acharya\,\orcidlink{0000-0002-9213-5329}\,$^{\rm 49}$, 
K.~Agarwal\,\orcidlink{0000-0001-5781-3393}\,$^{\rm II,}$$^{\rm 23}$, 
G.~Aglieri Rinella\,\orcidlink{0000-0002-9611-3696}\,$^{\rm 32}$, 
L.~Aglietta\,\orcidlink{0009-0003-0763-6802}\,$^{\rm 24}$, 
N.~Agrawal\,\orcidlink{0000-0003-0348-9836}\,$^{\rm 25}$, 
Z.~Ahammed\,\orcidlink{0000-0001-5241-7412}\,$^{\rm 132}$, 
S.~Ahmad\,\orcidlink{0000-0003-0497-5705}\,$^{\rm 15}$, 
I.~Ahuja\,\orcidlink{0000-0002-4417-1392}\,$^{\rm 36}$
% ...

\section*{Affiliation Notes}

$^{\rm I}$ Deceased\\
$^{\rm II}$ Also at: INFN Trieste\\
$^{\rm III}$ Also at: Max-Planck-Institut fur Physik, Munich, Germany\\
$^{\rm IV}$ Also at: Czech Technical University in Prague (CZ)\\
$^{\rm V}$ Also at: Instituto de Fisica da Universidade de Sao Paulo\\
$^{\rm VI}$ Also at: Dipartimento DET del Politecnico di Torino, Turin, Italy\\
$^{\rm VII}$ Also at: Department of Applied Physics, Aligarh Muslim University, Aligarh, India\\
$^{\rm VIII}$ Also at: Institute of Theoretical Physics, University of Wroclaw, Poland\\
$^{\rm IX}$ Also at: Facultad de Ciencias, Universidad Nacional Aut\'{o}noma de M\'{e}xico, Mexico City, Mexico\\

\section*{Collaboration Institutes}

$^{1}$ A.I. Alikhanyan National Science Laboratory (Yerevan Physics Institute) Foundation, Yerevan, Armenia\\
$^{2}$ AGH University of Krakow, Cracow, Poland\\
$^{3}$ Bogolyubov Institute for Theoretical Physics, National Academy of Sciences of Ukraine, Kyiv, Ukraine\\
$^{4}$ Bose Institute, Department of Physics  and Centre for Astroparticle Physics and Space Science (CAPSS), Kolkata, India\\
$^{5}$ California Polytechnic State University, San Luis Obispo, California, United States\\
$^{6}$ Central China Normal University, Wuhan, China\\
$^{7}$ Centro de Aplicaciones Tecnol\'{o}gicas y Desarrollo Nuclear (CEADEN), Havana, Cuba\\
$^{8}$ Centro de Investigaci\'{o}n y de Estudios Avanzados (CINVESTAV), Mexico City and M\'{e}rida, Mexico\\
$^{9}$ Chicago State University, Chicago, Illinois, United States\\
$^{10}$ China Nuclear Data Center, China Institute of Atomic Energy, Beijing, China\\
$^{11}$ China University of Geosciences, Wuhan, China\\
$^{12}$ Chungbuk National University, Cheongju, Republic of Korea\\
$^{13}$ Comenius University Bratislava, Faculty of Mathematics, Physics and Informatics, Bratislava, Slovak Republic\\
$^{14}$ Creighton University, Omaha, Nebraska, United States\\
$^{15}$ Department of Physics, Aligarh Muslim University, Aligarh, India\\
$^{16}$ Department of Physics, Pusan National University, Pusan, Republic of Korea\\
$^{17}$ Department of Physics, Sejong University, Seoul, Republic of Korea\\
$^{18}$ Department of Physics, University of California, Berkeley, California, United States\\
$^{19}$ Department of Physics, University of Oslo, Oslo, Norway\\
$^{20}$ Department of Physics and Technology, University of Bergen, Bergen, Norway\\
$^{21}$ Dipartimento di Fisica, Universit\`{a} di Pavia, Pavia, Italy\\
$^{22}$ Dipartimento di Fisica dell'Universit\`{a} and Sezione INFN, Cagliari, Italy\\
$^{23}$ Dipartimento di Fisica dell'Universit\`{a} and Sezione INFN, Trieste, Italy\\
$^{24}$ Dipartimento di Fisica dell'Universit\`{a} and Sezione INFN, Turin, Italy\\
$^{25}$ Dipartimento di Fisica e Astronomia dell'Universit\`{a} and Sezione INFN, Bologna, Italy\\
$^{26}$ Dipartimento di Fisica e Astronomia dell'Universit\`{a} and Sezione INFN, Catania, Italy\\
$^{27}$ Dipartimento di Fisica e Astronomia dell'Universit\`{a} and Sezione INFN, Padova, Italy\\
$^{28}$ Dipartimento di Fisica `E.R.~Caianiello' dell'Universit\`{a} and Gruppo Collegato INFN, Salerno, Italy\\
$^{29}$ Dipartimento DISAT del Politecnico and Sezione INFN, Turin, Italy\\
$^{30}$ Dipartimento di Scienze MIFT, Universit\`{a} di Messina, Messina, Italy\\
$^{31}$ Dipartimento Interateneo di Fisica `M.~Merlin' and Sezione INFN, Bari, Italy\\
$^{32}$ European Organization for Nuclear Research (CERN), Geneva, Switzerland\\
$^{33}$ Faculty of Electrical Engineering, Mechanical Engineering and Naval Architecture, University of Split, Split, Croatia\\
$^{34}$ Faculty of Nuclear Sciences and Physical Engineering, Czech Technical University in Prague, Prague, Czech Republic\\
$^{35}$ Faculty of Physics, Sofia University, Sofia, Bulgaria\\
$^{36}$ Faculty of Science, P.J.~\v{S}af\'{a}rik University, Ko\v{s}ice, Slovak Republic\\
$^{37}$ Faculty of Technology, Environmental and Social Sciences, Bergen, Norway\\
$^{38}$ Frankfurt Institute for Advanced Studies, Johann Wolfgang Goethe-Universit\"{a}t Frankfurt, Frankfurt, Germany\\
$^{39}$ Fudan University, Shanghai, China\\
$^{40}$ Gauhati University, Department of Physics, Guwahati, India\\
$^{41}$ Helmholtz-Institut f\"{u}r Strahlen- und Kernphysik, Rheinische Friedrich-Wilhelms-Universit\"{a}t Bonn, Bonn, Germany\\
$^{42}$ Helsinki Institute of Physics (HIP), Helsinki, Finland\\
$^{43}$ High Energy Physics Group,  Universidad Aut\'{o}noma de Puebla, Puebla, Mexico\\
$^{44}$ Horia Hulubei National Institute of Physics and Nuclear Engineering, Bucharest, Romania\\
$^{45}$ HUN-REN Wigner Research Centre for Physics, Budapest, Hungary\\
$^{46}$ Indian Institute of Technology Bombay (IIT), Mumbai, India\\
$^{47}$ Indian Institute of Technology Indore, Indore, India\\
$^{48}$ INFN, Laboratori Nazionali di Frascati, Frascati, Italy\\
$^{49}$ INFN, Sezione di Bari, Bari, Italy\\
$^{50}$ INFN, Sezione di Bologna, Bologna, Italy\\
$^{51}$ INFN, Sezione di Cagliari, Cagliari, Italy\\
$^{52}$ INFN, Sezione di Catania, Catania, Italy\\
$^{53}$ INFN, Sezione di Padova, Padova, Italy\\
$^{54}$ INFN, Sezione di Pavia, Pavia, Italy\\
$^{55}$ INFN, Sezione di Torino, Turin, Italy\\
$^{56}$ INFN, Sezione di Trieste, Trieste, Italy\\
$^{57}$ Inha University, Incheon, Republic of Korea\\
$^{58}$ Institute for Gravitational and Subatomic Physics (GRASP), Utrecht University/Nikhef, Utrecht, Netherlands\\
$^{59}$ Institute of Experimental Physics, Slovak Academy of Sciences, Ko\v{s}ice, Slovak Republic\\
$^{60}$ Institute of Physics, Homi Bhabha National Institute, Bhubaneswar, India\\
$^{61}$ Institute of Physics of the Czech Academy of Sciences, Prague, Czech Republic\\
$^{62}$ Institute of Space Science (ISS), Bucharest, Romania\\
$^{63}$ Institut f\"{u}r Kernphysik, Johann Wolfgang Goethe-Universit\"{a}t Frankfurt, Frankfurt, Germany\\
$^{64}$ Instituto de Ciencias Nucleares, Universidad Nacional Aut\'{o}noma de M\'{e}xico, Mexico City, Mexico\\
$^{65}$ Instituto de F\'{i}sica, Universidade Federal do Rio Grande do Sul (UFRGS), Porto Alegre, Brazil\\
$^{66}$ Instituto de F\'{\i}sica, Universidad Nacional Aut\'{o}noma de M\'{e}xico, Mexico City, Mexico\\
$^{67}$ iThemba LABS, National Research Foundation, Somerset West, South Africa\\
$^{68}$ Jeonbuk National University, Jeonju, Republic of Korea\\
$^{69}$ Korea Institute of Science and Technology Information, Daejeon, Republic of Korea\\
$^{70}$ Laboratoire de Physique Subatomique et de Cosmologie, Universit\'{e} Grenoble-Alpes, CNRS-IN2P3, Grenoble, France\\
$^{71}$ Lawrence Berkeley National Laboratory, Berkeley, California, United States\\
$^{72}$ Lund University Department of Physics, Division of Particle Physics, Lund, Sweden\\
$^{73}$ Marietta Blau Institute, Vienna, Austria\\
$^{74}$ Nagasaki Institute of Applied Science, Nagasaki, Japan\\
$^{75}$ Nara Women{'}s University (NWU), Nara, Japan\\
$^{76}$ National Centre for Nuclear Research, Warsaw, Poland\\
$^{77}$ National Institute of Science Education and Research, Homi Bhabha National Institute, Jatni, India\\
$^{78}$ National Nuclear Research Center, Baku, Azerbaijan\\
$^{79}$ National Research and Innovation Agency - BRIN, Jakarta, Indonesia\\
$^{80}$ Niels Bohr Institute, University of Copenhagen, Copenhagen, Denmark\\
$^{81}$ Nikhef, National institute for subatomic physics, Amsterdam, Netherlands\\
$^{82}$ Nuclear Physics Group, STFC Daresbury Laboratory, Daresbury, United Kingdom\\
$^{83}$ Nuclear Physics Institute of the Czech Academy of Sciences, Husinec-\v{R}e\v{z}, Czech Republic\\
$^{84}$ Oak Ridge National Laboratory, Oak Ridge, Tennessee, United States\\
$^{85}$ Ohio State University, Columbus, Ohio, United States\\
$^{86}$ Physics department, Faculty of science, University of Zagreb, Zagreb, Croatia\\
$^{87}$ Physics Department, Panjab University, Chandigarh, India\\
$^{88}$ Physics Department, University of Jammu, Jammu, India\\
$^{89}$ Physics Program and International Institute for Sustainability with Knotted Chiral Meta Matter (WPI-SKCM$^{2}$), Hiroshima University, Hiroshima, Japan\\
$^{90}$ Physikalisches Institut, Eberhard-Karls-Universit\"{a}t T\"{u}bingen, T\"{u}bingen, Germany\\
$^{91}$ Physikalisches Institut, Ruprecht-Karls-Universit\"{a}t Heidelberg, Heidelberg, Germany\\
$^{92}$ Physik Department, Technische Universit\"{a}t M\"{u}nchen, Munich, Germany\\
$^{93}$ Politecnico di Bari and Sezione INFN, Bari, Italy\\
$^{94}$ Research Division and ExtreMe Matter Institute EMMI, GSI Helmholtzzentrum f\"ur Schwerionenforschung GmbH, Darmstadt, Germany\\
$^{95}$ Saga University, Saga, Japan\\
$^{96}$ Saha Institute of Nuclear Physics, Homi Bhabha National Institute, Kolkata, India\\
$^{97}$ School of Physics and Astronomy, University of Birmingham, Birmingham, United Kingdom\\
$^{98}$ Secci\'{o}n F\'{\i}sica, Departamento de Ciencias, Pontificia Universidad Cat\'{o}lica del Per\'{u}, Lima, Peru\\
$^{99}$ SUBATECH, IMT Atlantique, Nantes Universit\'{e}, CNRS-IN2P3, Nantes, France\\
$^{100}$ Sungkyunkwan University, Suwon City, Republic of Korea\\
$^{101}$ Suranaree University of Technology, Nakhon Ratchasima, Thailand\\
$^{102}$ Technical University of Ko\v{s}ice, Ko\v{s}ice, Slovak Republic\\
$^{103}$ The Henryk Niewodniczanski Institute of Nuclear Physics, Polish Academy of Sciences, Cracow, Poland\\
$^{104}$ The University of Texas at Austin, Austin, Texas, United States\\
$^{105}$ Universidad Aut\'{o}noma de Sinaloa, Culiac\'{a}n, Mexico\\
$^{106}$ Universidade de S\~{a}o Paulo (USP), S\~{a}o Paulo, Brazil\\
$^{107}$ Universidade Estadual de Campinas (UNICAMP), Campinas, Brazil\\
$^{108}$ Universidade Federal do ABC, Santo Andre, Brazil\\
$^{109}$ Universitatea Nationala de Stiinta si Tehnologie Politehnica Bucuresti, Bucharest, Romania\\
$^{110}$ University of Cape Town, Cape Town, South Africa\\
$^{111}$ University of Derby, Derby, United Kingdom\\
$^{112}$ University of Houston, Houston, Texas, United States\\
$^{113}$ University of Jyv\"{a}skyl\"{a}, Jyv\"{a}skyl\"{a}, Finland\\
$^{114}$ University of Kansas, Lawrence, Kansas, United States\\
$^{115}$ University of Liverpool, Liverpool, United Kingdom\\
$^{116}$ University of Science and Technology of China, Hefei, China\\
$^{117}$ University of Silesia in Katowice, Katowice, Poland\\
$^{118}$ University of South-Eastern Norway, Kongsberg, Norway\\
$^{119}$ University of Tennessee, Knoxville, Tennessee, United States\\
$^{120}$ University of the Witwatersrand, Johannesburg, South Africa\\
$^{121}$ University of Tokyo, Tokyo, Japan\\
$^{122}$ University of Tsukuba, Tsukuba, Japan\\
$^{123}$ Universit\"{a}t M\"{u}nster, Institut f\"{u}r Kernphysik, M\"{u}nster, Germany\\
$^{124}$ Universit\'{e} Clermont Auvergne, CNRS/IN2P3, LPC, Clermont-Ferrand, France\\
$^{125}$ Universit\'{e} de Lyon, CNRS/IN2P3, Institut de Physique des 2 Infinis de Lyon, Lyon, France\\
$^{126}$ Universit\'{e} de Strasbourg, CNRS, IPHC UMR 7178, F-67000 Strasbourg, France, Strasbourg, France\\
$^{127}$ Universit\'{e} Paris-Saclay, Centre d'Etudes de Saclay (CEA), IRFU, D\'{e}partment de Physique Nucl\'{e}aire (DPhN), Saclay, France\\
$^{128}$ Universit\'{e}  Paris-Saclay, CNRS/IN2P3, IJCLab, Orsay, France\\
$^{129}$ Universit\`{a} degli Studi di Foggia, Foggia, Italy\\
$^{130}$ Universit\`{a} del Piemonte Orientale, Vercelli, Italy\\
$^{131}$ Universit\`{a} di Brescia, Brescia, Italy\\
$^{132}$ Variable Energy Cyclotron Centre, Homi Bhabha National Institute, Kolkata, India\\
$^{133}$ Warsaw University of Technology, Warsaw, Poland\\
$^{134}$ Wayne State University, Detroit, Michigan, United States\\
$^{135}$ Yale University, New Haven, Connecticut, United States\\
$^{136}$ Yildiz Technical University, Istanbul, Turkey\\
$^{137}$ Yonsei University, Seoul, Republic of Korea\\
$^{138}$ Affiliated with an institute formerly covered by a cooperation agreement with CERN\\
$^{139}$ Affiliated with an international laboratory covered by a cooperation agreement with CERN.\\

\end{flushleft} 
  
\end{document}